\documentclass[aps,a4paper,superscriptaddress,onecolumn,showpacs,preprintnumbers,amsmath,amssymb]{revtex4}
\usepackage{graphicx}
\usepackage{bm}

\usepackage{amsmath}
\usepackage{amssymb}
\usepackage{graphicx}
\usepackage{gensymb}
\usepackage{color}
\usepackage{ulem}

\begin{document}

\title{Neutron Interferometry constrains dark energy chameleon fields}

\newcommand{\ipht}{Institut de Physique Th\'eorique, CEA, IPhT, CNRS, URA 2306, F-91191 Gif / Yvette Cedex, France}
\newcommand{\ill}{Institut Laue-Langevin, 38042 Grenoble, France}
\newcommand{\ati}{Technische Universit{\"a}t Wien, Atominstitut, 1020 Wien, Austria}
\newcommand{\lpsc}{LPSC, Universit\'e Grenoble Alpes, CNRS/IN2P3 F-38026 Grenoble, France}

\author{H. Lemmel}
\affiliation{\ati,\ill}

\author{Ph.~Brax}
\affiliation{\ipht}

\author{A.~N.~Ivanov}
\affiliation{\ati}

\author{T.~Jenke}
\affiliation{\ati}

\author{G.~Pignol}
\affiliation{\lpsc}

\author{M.~Pitschmann}
\affiliation{\ati}

\author{T.~Potocar}
\affiliation{\ati}

\author{M.~Wellenzohn}
\affiliation{\ati}

\author{M.~Zawisky}
\affiliation{\ati}

\author{H.~Abele}
\affiliation{\ati}

\date{\today}

\begin{abstract}
We present phase shift measurements for neutron matter waves in vacuum
and in low pressure Helium using a method originally developed for
neutron scattering length measurements in neutron interferometry. We
search for phase shifts associated with a coupling to scalar
fields. We set stringent limits for a scalar chameleon field, a
prominent quintessence dark energy candidate. We find that the
coupling constant $\beta$ is less than 1.9 $\times10^7$~for $n=1$
at 95\% confidence level, where $n$ is an input parameter of the
self--interaction of the chameleon field $\varphi$ inversely
proportional to $\varphi^n$.
\end{abstract}


%
\maketitle

\section{Introduction}
\label{intro}
The accelerating expansion of the universe suggests that most of the
energy in the universe is 'dark energy'. The nature and origin of this
energy remain unknown.
Candidates for dark energy are either Einstein's cosmological constant or dynamical dark energy,
i.e. the so-called quintessence canonical scalar field $\varphi$,
responsible for the late--time acceleration of the universe expansion.
Chameleon fields are a prime example of dynamical dark energy.
Their effective mass depends on the energy density of
matter in which it is immersed [1]. As a result, in a sufficiently dense
environment the chameleon field is very massive and, correspondingly,
substantially Yukawa--suppressed, i.e. very short--ranged. In turn,
it is essentially massless on cosmological scales [2, 3].
Because of its sensitivity on the environment, such a mass-changing
scalar field has been called {\it chameleon}. Moreover, the chameleon field always couples
to  matter and generates a fifth force with an effective
range inversely proportional to its effective mass.

All models of dark energy involve a light scalar field \cite{ Copeland:2006wr,Clifton:2011jh} whose effects on solar system tests of gravity needs to be shielded. Three main screening mechanisms \cite{Joyce:2014kja} have been unraveled so far. The K--mouflage and Vainshtein screenings are very powerful inside a large domain surrounding the earth, rendering their test in laboratory experiments extremely arduous. On the other hand, the chameleon mechanism is at work in the presence of dense objects and can be tested in near-vacuum experiments \cite{Brax:2014zta}. This is the case for the Eotwash \cite{Upadhye:2012qu} and Casimir experiments \cite{Brax:2010xx}, where the boundary plates are screened. Another way of testing the chameleon mechanism involves small and unscreened objects, like neutrons under certain conditions~\cite{Brax2013}.

Concerning chameleon models, a chameleon-photon coupling $g_{\rm eff} = \beta_{\gamma}/M_{\rm Pl}$
has been proposed, and the detailed analysis of the chameleon--photon interaction
and a comparison with the cosmological data has been carried out in
\cite{Brax2007,Brax2007a,Burrage2008,Burrage2009,Davis2009}.
A search for photon--chameleon--photon transition has been performed by the experiment CHASE (the GammeV CHameleon Afterglow SEarch)\cite{Chou2009} and by the Axion Dark Matter eXperiment (ADMX) \cite{Rybka2010}. A search for chameleon particles created via photon-chameleon oscillations
within a magnetic field is described in~\cite{Steffen10}.

Searches with neutrons directly test the chameleon-matter interaction $\beta$ and do not rely on the existence of a chameleon--photon--interaction. The coupling $\beta$ is restricted from below, e.g. $\beta$ must be larger than 50 at $n$ = 1~\cite{Brax2011}, and experiments with neutrons have the potential ultimately to find a chameleon field or exclude it in the whole parameter space.

As it has been pointed out by Pokotilovski~\cite{Pokotilovski2013}, the
use of a neutron Lloyd's interferometer for measurements of the
phase--shift of the wave function of cold neutrons should allow to
determine the chameleon--matter coupling constant.
The \textit{q}\textsc{Bounce} collaboration has searched for the
chameleon field using gravity resonance spectroscopy and ultra-cold
neutrons \cite{Jenke2011,Abele2009,Jenke2009,Abele2010}. In a recent
experiment \cite{Jenke2014}, the upper limit for $\beta$ has been
determined as $\beta < 5.8 \times 10^{8}$ which is five orders of
magnitude below the previous limit determined by atomic spectra
\cite{Brax2011}.

Here we present a new search for chameleon fields by means of neutron
interferometry as proposed in \cite{Brax2013}.
The self--interaction of the chameleon field $\varphi$ and its
interaction to an environment with mass density $\rho$ are described
by the effective potential \cite{Khoury2004prl,Mota2007}
\begin{equation}
\label{eq:1}
	\mathcal{V}_{\rm eff}(\varphi) =
        \frac{\Lambda^{n+4}}{\varphi^n} + \frac{\beta \, \rho\,
          \hbar^3 c^3 \varphi}{ M_{\rm Pl}},
\end{equation}
where $\beta$ is the coupling constant, $n$ is an input parameter (the
so--called Ratra--Peebles index) and $\Lambda \approx 2.4\times
10^{-12}$ GeV defines an energy scale \cite{Brax2013}.  $M_{\rm
  Pl}=\sqrt{\hbar c/(8\pi G)} = 4.341\times 10^{-9}$ kg denotes the
reduced Planck mass.
The chameleon field
$\varphi$ creates a potential for neutrons given by $V=\beta \,
\varphi \, m/M_{\rm Pl}$ where $m$ denotes the neutron mass. When
passing this potential, neutrons accumulate the phase
\begin{equation}\label{field2phase}
  \zeta = - \frac{m}{k \hbar^2 }\; \int V(x)\, dx = - \frac{m}{k
    \hbar^2 }\; \int \beta \frac m {M_{\rm Pl}} \, \varphi(x)\,
  dx,
\end{equation}
where $k$ denotes the neutron wave vector modulus $k=2\pi/\lambda$.

For strong coupling ($\beta \gg 1$) the chameleon field is suppressed
at the presence of matter, even at low mass densities like air at
ambient pressure. Only in vacuum the chameleon field can persist. By
placing a vacuum cell into one arm of the neutron interferometer and
allowing ambient air in the other arm we can directly probe the
chameleon field. The setup resembles a standard setup for measuring
neutron scattering lengths \cite{rauch-werner}, but instead of
measuring the phase shift of sample material we measure the phase
shift of vacuum.

The chameleon field vanishes at the walls of the vacuum chamber but
increases bubble-like towards the middle of the chamber,
cf. Fig. \ref{figsetup} (c). The more the field increases the lower
the remaining gas pressure is, i.e. the better the vacuum is. Thus we
have two options of performing a relative phase measurement which is
necessary to cancel the unknown intrinsic interferometer phase and the
air phase shift. In the pressure mode we vary the pressure in the
vacuum cell by letting in different amounts of Helium. In the profile
mode we keep the pressure constant but move the chamber transversally
to the beam in order to record a profile of the chameleon
bubble. Neither method detects any chameleon-like signature,
giving rise to new constraints of the chameleon theory.


\section{Setup}

\begin{figure}[htb]
\centering\includegraphics[scale=0.4]{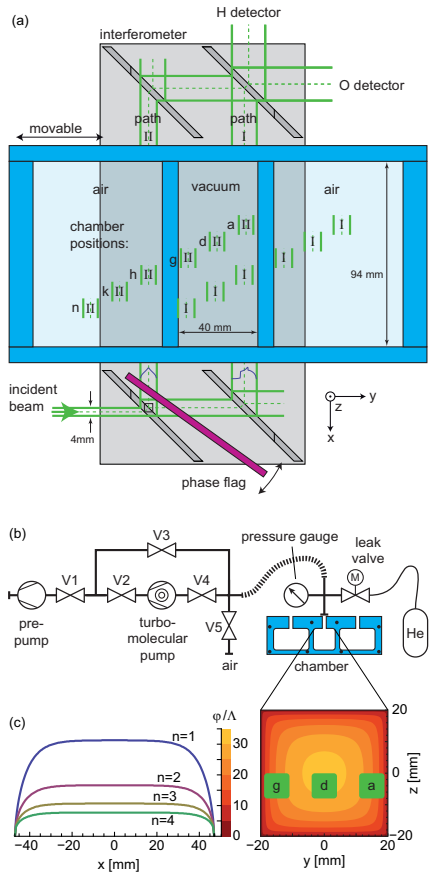}
\caption{(a) Top view of the interferometry setup shown in chamber position 'h'.
The chamber box (blue) can be moved transversally allowing the beams to pass at
different positions, labeled by 'a' to 'n'. (b) Scheme of the vacuum
handling and axial view of the vacuum chamber.
(c) Longitudinal and transverse bubble shape of the chameleon field
in the vacuum cell. The beam positions 'a', 'd' and 'g' are indicated by green rectangles.}
\label{figsetup}
\end{figure}

The experiment is carried out at the neutron interferometry setup S18
at the Institut Laue-Langevin (ILL) in Grenoble.  A perfect crystal
silicon interferometer is used, Fig. \ref{figsetup} (a), at
45$\degree$ Bragg angle and $2.72$ \AA$ $ mean wave length $\lambda$
with $0.043$ \AA$ $ wavelength distribution width (FWHM). The two beam
paths within the interferometer are separated by 50 mm over a length
of 160 mm. Neutron detectors with an efficiency above 99\% measure the
intensities of the two exit beams labeled O and H respectively.  A
vacuum chamber with inner dimensions 40 x 40 x 94 mm is inserted in
the left or right beam path. The other beam path always contains one
of the two air chambers which sit alongside the vacuum chamber. The
whole chamber box can be moved sidewards for swapping the vacuum cell
between the left and the right beam path and to probe different beam
trajectories within the vacuum cell. The air chambers ensure that both
beam paths contain the same amount of wall material (aluminium). In
addition, the extension of the vacuum cell by air chambers minimizes
possible disturbances of the thermal environment of the crystal when
the chamber box is moved. We label different chamber positions by the
letters 'a' to 'n' as indicated in the figure.

The air chambers are connected to ambient air by a hole in the top of
the chambers. The vacuum chamber is connected to a vacuum control
system consisting of pressure gauge, motorized leak valve and pumps,
as indicated in Fig. \ref{figsetup} (b). The pumps (pre-pump and
turbomolecular pump) are running continuously while a controlled
amount of Helium is let in through the leak valve in order to tune the
pressure. The pressure gauge is corrected for the use with Helium.

\section{Data acquisition and evaluation}

\begin{figure}
\centering\includegraphics[scale=0.65]{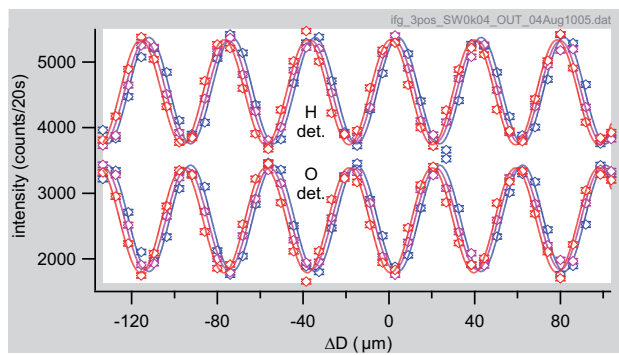}
\caption{Recorded intensity oscillations between O and H detector as a
  function of the optical path length difference $\Delta D$ created by
  rotating the phase flag. The three curves in red, purple and blue
  represent the interferograms at the 'a', 'd' and 'g' position
  respectively. The phase shift between these raw curves is created by
  position dependent wall thickness variations,
  cf. Fig. \ref{figprofile}. }
\label{figifg}
\end{figure}

\begin{figure}
\centering\includegraphics[scale=0.6]{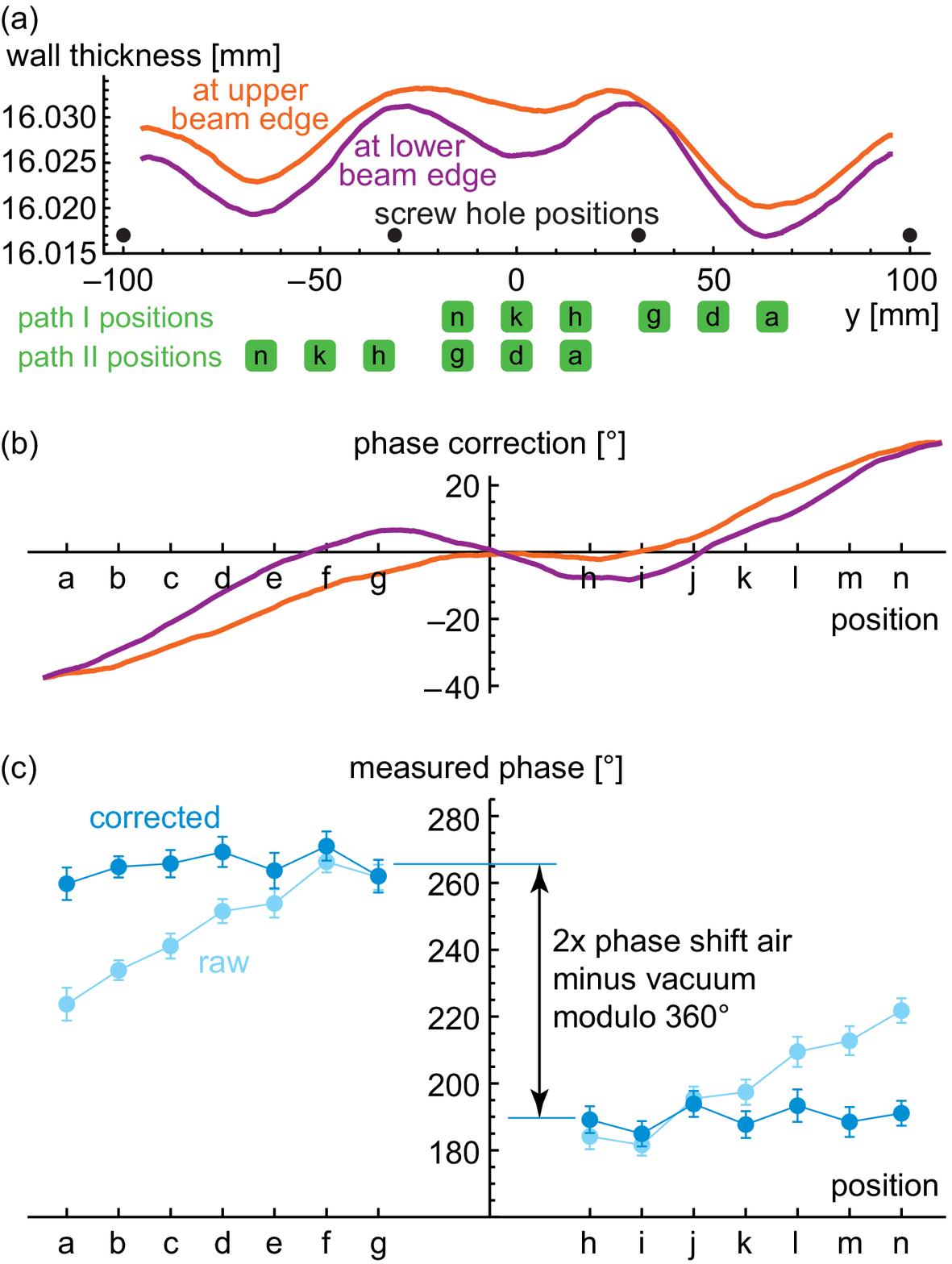}
\caption{(a) Mechanical measurement: measured total thickness of the entry and exit walls at
two different vertical positions. The thickness is increased around
the screw holes (black dots). The positions of beam path I and II are indicated
for several chamber positions. (b) Phase shift calculated from (a), caused by the different
amount of wall material (aluminium) in path I and II for different chamber positions.
(c) Phase profile of the vacuum chamber with and without correction for the wall thickness.
On the left side (a-g) path I passes air and path II passes vacuum; vice versa on the right side. }
\label{figprofile}
\end{figure}

Phases in neutron interferometry are measured by rotating an auxiliary
phase flag and recording the intensity oscillations between O and H
detector, cf. Fig. \ref{figifg}. Such interferograms are measured
before and after some parameter change. The shift of the sine curves
with respect to each other represents the phase shift induced by the
parameter change. The recording of each interferogram takes typically
half an hour, and during that time the intrinsic phase of the
interferometer can drift due to temperature changes or other
environmental factors. To compensate such drifts we interlace phase
flag movement and parameter change. The phase flag is rotated to the
first angular position and neutrons are counted for a certain amount
of time for each parameter setting. Then the phase flag is rotated to
the next position and neutrons are counted again for all parameter
settings etc. In the end we obtain interferograms measured
simultaneously for all parameter settings. Their relative phases are
free of phase drifts.

We use the largest neutron interferometer available~\cite{Zawisky2002}
with a loop size of 50 x 160 mm in order to maximize the size of the
vacuum cell. Such big single crystal interferometers are extremely
sensitive to temperature gradients, air flow, vibrations, bending,
etc. Hence the interference contrast (fringe visibility) is restricted
to about 10\% to 30\%. The interferograms look a bit more noisy than
what can be explained by pure counting statistics. This means that the
phase is slightly fluctuating within the recording time of each
interferogram. We conservatively account for this noise by performing
a $\chi^2$ test for each sine fit and by blowing up the fit error (by
a factor of about 2) such that the $\chi^2$ test is satisfied.

\begin{figure}
\centering\includegraphics[scale=0.6]{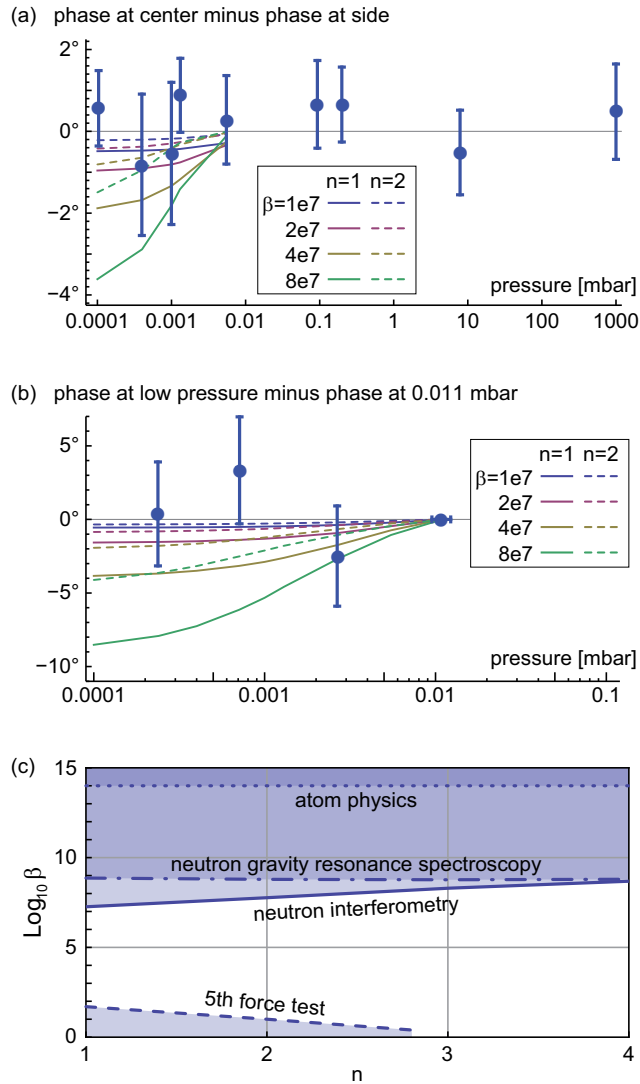}
\caption{ Measured phase shifts in the profile mode (a) and pressure
  mode (b) compared to calculations for different values of $\beta$
  and $n$. (c) Exclusion plot comparing our results with other
  experiments. The limit for $\beta$ at 95\% confidence level is shown
  for different values of $n$.  }
\label{figresult}
\end{figure}

\subsection{Profile mode}

In the profile mode we measure at up to 14 transverse beam positions
for each phase flag position in order to look for bubble-like phase
profiles. Ideally, the entry and exit walls of the vacuum chamber are
flat and parallel and therefore would not alter the phase if the
chamber is transversally moved. Unfortunately, the screw holes of our
walls have been drilled after the surfaces had been polished. As a
consequence, the surfaces are elevated by a few microns around each
screw hole, and all our data in profile mode require a position
dependent phase correction, based on a careful mapping of the wall
thickness, as indicated in Fig. \ref{figprofile} (a) and (b).
The screw hole positions are indicated by black dots in
Fig. \ref{figprofile} (a) and Fig. \ref{figsetup} (b).
Fortunately, the correction depends mainly linear on the beam
position, while the shape of the chameleon bubble is expected to be
mainly parabolic. To be precise, the chameleon profile must be
symmetric with respect to the cell center, and can therefore consist
only of quadratic and higher even orders. Thus, there is no danger
that the wall thickness correction completely mimics or hides the
chameleon feature. We determine the thickness correction at the upper
and the lower edge of the beam, and use the average as correction and
a quarter of the difference as uncertainty of the correction.

Fig. \ref{figprofile} (c) shows the recorded phase over a complete profile. In position 'a' to 'g' path I passes air and path II passes vacuum while in position 'h' to 'n' it is the other way round, cf. Fig. \ref{figsetup} (a). The slope within each group is caused by the thickness variation of the chamber walls. The step between the two groups comes from the sign change of the air phase shift when air and vacuum are swapped between the beam paths.

The height of the chameleon bubble can be determined by comparing the
phase at the center of the chamber with the phase at its side, close
to the chamber walls. Therefore we make most of our measurements at
positions 'a', 'd' and 'g'. Fig. \ref{figresult} (a) summarizes the
result of the bubble height measurements for various pressure
settings. The statistical error of the phase can be reduced to
typically 0.9$\degree$ by averaging over 15 measurements. However, the
thickness correction, which is applied after the statistical
averaging, increases the error again to typically 2.5$\degree$.

\subsection{Pressure mode}

In the pressure mode we apply four different pressures at each phase
flag position. A quick pressure change is only possible in the
pressure range of the turbomolecular pump, i.e. below $10^{-2}$
mbar. The average of four such runs is shown in Fig. \ref{figresult}
(b). In order to compensate phase drifts we use the phase at the
highest pressure (0.011 mbar) as reference and determine the phase
shift between this pressure and the other pressure values. The
magnitude of the phase shift created by the Helium itself is in the
order of $\lesssim 0.001\degree$ in this pressure range and can be
neglected.

\section{Limit calculation}

The solution of the chameleon field in vacuum confined between two
walls at $x=\pm d/2$ is given analytically \cite{Ivanov2013} by
\begin{equation}\label{bubble1D}
 \varphi_{\rm 1D}(x) = \Lambda \left\{ \frac{\Lambda d}{\hbar c} \,
 \frac{n\!+\!2}{2\sqrt 2} \left[ 1 - \left(\frac x d\right)^2
   \right]\right\}^{\frac 2{n+2}}.
\end{equation}
For higher dimensions and for finite gas pressure the calculation has
to be done numerically. The left side of Fig. \ref{figsetup} (c) shows
the longitudinal field profile $\varphi_{\rm 3D}(x,y\!=\!0,z\!=\!0)$
along the center of the chamber calculated in 3D and for vacuum. It
vanishes at the walls and increases towards the middle. Over most of
the range it is nearly constant because it is limited by the much
narrower transverse confinement. The transverse field distribution
$\varphi_{\rm 2D}(y,z) \approx \varphi_{\rm 3D}(0,y,z)$ is shown on
the right side.

Since the full 3D calculation is very time consuming we assume to good
approximation that the field depends only on the transverse
coordinates. We account for the longitudinal drop close to the walls
by calculating an effective chamber length $l_{\rm eff}$ such that
\begin{equation}
  \int_{-l/2}^{l/2} \varphi_{\rm 3D}(x,0,0)\, dx = \varphi_{\rm
    3D}(0,0,0) \; l_{\rm eff}.
\end{equation}
Thus the true length $l=94$ mm reduces effectively to $l_{\rm
  eff}=\{84,85.6,86.8,87.6\}$ mm respectively for $n=\{1,2,3,4\}$. The
expected phase shift $\zeta$ given by Eq. (\ref{field2phase})
simplifies to
\begin{equation}
  \zeta = \frac{m}{k \hbar^2}\,\beta\,\frac{m}{ M_{\rm Pl}}\;  \varphi_{\rm 2D}(y,z) \; l_{\rm eff}
\end{equation}
and is plotted in Fig. \ref{figresult} (a) and (b) assuming various
values for $\beta$ and $n$.

We calculate limits for $\beta$ by comparing the calculated phase
shifts $\xi$ with the measured phase shifts $\zeta \pm \sigma$. We
assume certain values of $\beta$ and calculate the corresponding
probability $p$.
\begin{eqnarray}
p(\beta) &=& \frac{ \exp \left[- \frac 1 2
    \, \chi(\beta)^2 \right] }{ \int_0^{\beta_{\rm max}} \exp \left[-
    \frac 1 2 \, \chi(\beta)^2 \right] \, d\beta } \\ \chi(\beta)^2 &=&
\sum_i \frac {\left[\xi(\beta)_i-\zeta_i \right]^2}{\sigma_i^2}
\end{eqnarray}
The sum goes over the data points shown in Fig. \ref{figresult} (a)
and (b).  We determine the limit $\beta_{\rm lim}$ with 95\%
confidence level by numerically solving the equation
\begin{equation}
  \int_0^{\beta_{\rm lim}} p(\beta) \, d\beta = 95\%.
\end{equation}
The calculation is repeated for different values of $n$ yielding the
following results.
\begin{eqnarray} \label{limits}
  \beta_{\rm lim} = \left\{\begin{array}{r@{\quad,\quad}l} 1.9 \times 10^7 & n=1\,, \\ 5.8
    \times 10^7 & n=2\,, \\ 2.0 \times 10^8 & n=3\,, \\ 4.8 \times 10^8 &
    n=4 \,.\end{array}\right.
\end{eqnarray}
Both the profile mode and the pressure mode contribute about equally
to these limits. For large $n$ the profile mode becomes less sensitive
because the bubble shape becomes flatter on the top.


\section{Conclusion}

Our search for chameleons by means of neutron interferometry failed
in finding ones but succeeded in deriving new upper bounds for the
coupling constant $\beta$, listed in Eq. (\ref{limits}). For $n=1$
the new limit is a factor of 30 below the previous one which has been
obtained by gravity resonance spectroscopy, cf. Fig. \ref{figresult}
(c). There remains a range of five orders of magnitude for $\beta$
where chameleons have not been excluded yet.


\section{Acknowledgements}

We thank Helmut Rauch and Martin Suda for useful discussions and
Manfried Faber for theoretical support.

We gratefully acknowledge support from the Austrian Science Fonds (FWF) under Contracts
No. I529-N20, I530-N20, I531-N20,  I689-N16, and I862-N20.

\textit{Note added - During the preparation of this Letter, another interfometric experiment searching for the chameleon, using ultracold Cs atoms, has been reported \cite{atom2015}.}

\bibliography{chameleon_arXiv}

\end{document}